\documentstyle[aps,prl,multicol,epsf]{revtex}
\begin{document}
\newcommand{\bra}{\langle} 
\newcommand{\ket}{\rangle}
\newcommand{\be}{\begin{equation}}
\newcommand{\ee}{\end{equation}} 
\newcommand{\del}{\partial}
\newcommand{\E}{\mbox{erf}}
\title{Statistical properties of spike trains: universal and
stimulus-dependent
aspects}
\author{Naama Brenner$^1$, Oded Agam$^2$, William Bialek$^1$ 
and Rob de Ruyter van
Steveninck$^1$}
\address{$^1$ NEC Research Institute, 4 Independence Way, Princeton, NJ
08540\\
$^2$ Racah Institute of Phsyics, Hebrew University, Jerusalem}
\maketitle
\begin{abstract}
Statistical properties of spike trains measured from a sensory neuron 
{\em in-vivo} are studied experimentally and theoretically. Experiments
are performed on an identified neuron in the visual system of the blowfly.
It is shown
that the spike trains exhibit universal behavior over short time, 
modulated by a stimulus-dependent
envelope over long time.  A model of the neuron as a nonlinear 
oscillator driven by noise and an external stimulus, 
is suggested to account for these results.
The model enables a theoretic distinction of the effects 
of internal neuronal properties 
from effects of external stimulus properties, and their identification in the
measured spike trains.
The universal regime is characterized by one dimensionless parameter,
representing the internal degree of irregularity, which is determined both by
the sensitivity of the neuron and by the
properties of the noise.  The envelope
is related in a simple way to properties of the input stimulus as seen through
nonlinearity of the neural response.  Explicit formulas are
derived for different statistical properties in both the universal and 
the stimulus-dependent regimes. These formulas are in very good agreement
with the data in both regimes.
\end{abstract}
\begin{multicols}{2}
\section{Introduction} 

Many cells in the nervous system respond to stimulation by generating
action potentials (spikes). 
Time sequences of these spikes are the basis
for encoding information and for communication between neurons
\cite{Spikes97}.
A pattern of spikes across time
contains, in addition to the message being encoded, the signature
of the biophysical spike generation mechanism, and of 
the noise in the neuron and its environment \cite{Jack75}. 
These
factors are generally inter-related in a complicated way, and it is not
clear how to disentangle their effect on the measured spike train.

The biophysical mechanism for generating action potentials was first
described successfully by Hodgkin and Huxley \cite{HH}. Their
description accounted for the stereotyped shape of an action potential,
which is a robust property, largely independent of external conditions.
The Hodgkin Huxley model describes
the neuron as a complex dynamical system; sustained firing
(a continuous train of spikes) comes about when the dynamical system is
driven into an oscillatory mode.
This picture is consistent with
experiments on isolated neurons: many of these tend to fire
periodic spike trains in response to direct current injection, 
implying an oscillator like behavior. The
frequency of these trains is deterministically related to the strength
of the applied current; Adrian (1928) suggested long ago
that this property
could be used to code the strength of the input.
Different neurons vary in the shape of the response function relating
frequency to input.  Following
Hodgkin and Huxley, many microscopic models of the neuron 
were constructed in the same spirit \cite{Jack75}.
One aim of this type of modeling is to produce the
different frequency-current ($f/I$) response curves by fitting
model parameters.

Spike trains measured {\em in vivo}, 
however, show a very different behavior: many
neurons seem to fire stochastically, 
even when external conditions are held fixed. This fact initiated
what seems to be an unrelated line of research, that of describing 
spike trains by models of stochastic processes \cite{Holden76,Sampath77}. 
These models can sometimes describe correctly statistical properties 
of the spikes trains, such as
the distribution of intervals, but in general the parameters of the
models remain unrelated to physiological characteristics
of real systems \cite{Tuckwell88}.

Several fundamental questions concerning the
statistical properties of spike trains thus remain unresolved, 
despite the large 
literature on this subject:
How is the periodic behavior of the isolated neuron
to be reconciled with the more irregular
behavior in a complex network? What is a useful characterization of the
degree of this irregularity, and how does it depend on external
conditions? How sensitive are the statistical properties
to the microscopic biophysical details of the neuron, 
and to the statistics of the noise? Can effects of the sensory stimulus
be separated and recognized at the output?

Here we present a theory which provides some 
answers to the above questions. We
use the notion of a frequency function to describe the neuron's
response \cite{Gutkin98}, and connect it to the
stochastic firing in a network through the introduction of noise.  
Under some conditions we find that
the statistical properties of spike trains are {\it
universal} on the time scale of a few spikes \cite{Brenner98}.  
This means that they are
independent of the details of the internal oscillator,
of the noise and of external stimulation.
All these are captured
by a single dimensionless parameter, related to the phase diffusion 
of the oscillator; this parameter characterizes the internal
irregularity of the point process. On the time scale of many spikes, the
universal behavior is modulated by an envelope reflecting the input
stimulus. We present experimental data for the statistical properties of
spikes trains measured from an identified motion sensor in the visual
system of the blowfly, under various external stimulation. These
data are shown to be very well described by the theory.

The paper is organized as follows: In section 2 we define our model, and 
show how the frequency of the oscillator is related to the rate
of the measured point process.
In Section 3 we consider the statistical properties of the model,
when the fluctuations in the 
inputs are rapid relative to the typical interspike time $1/r$. 
We derive explicit formulas for the statistical properties of the spike
train, and show that all details affect these properties only through the
irregularity parameter.
In Section 4, we consider the case of an additional slow time
scale in the inputs, which is much longer than $1/r$. We show that
the conditional rate can be approximated by
a product of two distinct parts: a universal part,
depending only on the irregularity parameter, and a
stimulus-dependent part, which modulates it.  In each section a
comparison of the theoretic results with measurements from the fly is
presented.

\section{Frequency Integrator}

A sequence of spikes will be described as a train of
Dirac $\delta$-functions:
\be
\rho(t) ~=~ \sum_{k} \delta(t-t_k).
\ee
In this approximation the height and shape of the
action potentials are neglected, and all the information 
is contained in their arrival times: the spike train is a point
process. 
If the system is driven by a signal $s(t)$ and a noise $n(t)$,
both continuous functions of time, then we can imagine that the neuron
evaluates some functional  ${\cal F}[s(t),n(t)]$ and produces a spike
when this crosses a threshold:

\begin{equation}
\rho(t) = \sum_k \delta\Bigg({\cal F}[s(t'),n(t')] - k \Bigg) 
~\frac{d{\cal F}}{dt}.  \label{rho}
\end{equation}
Formula (\ref{rho}) describes a very general class of models:
a spike is generated
when ${\cal F}$ crosses a fixed threshold,
and the process resets after spiking.
The operator ${\cal F}$
can be linear or nonlinear,
deterministic or stochastic, and can depend on the history of
the signal and the noise in a complicated way \cite{Bruckstein79}.
We focus on the following more specific form of ${\cal F}$:

\begin{equation}
{\cal F}[s(t),n(t)] = ~ \int
_0
^t 
f\!\left[s(u)+n(u)\right] ~du ~\equiv~\Phi(t),   \label{F}
\end{equation}

\noindent
where $f\geq 0$ is the frequency response function characterizing the
neuron. This model is related to Integral Frequency Pulse Modulation
models and to the standard integrate-and-fire model
\cite{Gestri71,Bruckstein79,Tuckwell88}. 
The motivation for defining a deterministic
frequency response function comes from the measured behavior of 
isolated neurons in response to direct current injection.
When driven by a constant
stimulus $s$, in the absence of noise,
our model neuron generates a periodic spike train
with a frequency $f(s)$, consistent with the behavior in isolation. 
Starting from a microscopic level of modeling, 
many parameters need to be tuned to produce a required form of the
$f/I$
relation \cite{Troyer97}. Here we use the $f/I$ relation as a 
phenomenological description of the neuron, and base the statistical
theory on this description. 

Now we would like to ``embed'' our model neuron in a noisy environment,
such as a complex sensory network, while it is still subject to a
constant stimulus $s$. In general there can be many
noise sources in such a network: the signals coming in from the external
world are not perfect, the sensory apparatus (such as photoreceptors)
is noisy,
connections between cells in the network introduce noise, and finally
the cell itself can generate noise (for example, channel noise
\cite{Schneidman98}).
We introduce noise as an additional random function $n(t)$ added
to the input. This simplified scheme is justified by the fact that
final results do not depend on the details of the noise distribution,
therefore $n(t)$ is understood as an effective noise.

Retaining the notion of a local frequency, the noise
causes frequency modulations in the spike train. Under the
conditions  $s\!=\!{\rm const}, n\!=\!0$ the oscillatory behavior
is related to some periodic trajectory in parameter space. Assuming
that this trajectory is
stable, the addition of $n(t)$ will cause the system to occupy a volume
in parameter space surrounding this trajectory. The phase of the oscillator
will not advance at a constant rate $f(s)$, but instead will be given by

\be
\dot {\Phi}(t)= f[s\!+\!n(t)].
\ee

\noindent The strength of the noise
and the sensitivity of $f(\cdot)$ to changes in the inputs, both determine
the frequency modulation depth, or the amount of randomness in the phase
advancement. 

The frequency function is an internal property of the neuron. One
would like to
relate it to the firing rate function in the presence
of noise, which can be measured experimentally. Considering still
the case of a
constant $s$ and 
introducing the average over noise
$\bra\cdots\ket$, the average spike train is

\be
\bra\rho_s(t)\ket ~=~ \sum_k \Bigg\langle\delta\left(
\Phi(t)-k\right)~ 
\dot{\Phi}(t)\Bigg\rangle.
\ee
Using the Poisson summation
formula, 
\be
\sum_k \delta(x-k) = \sum_m e^{i2\pi mx}
\ee
we can write the average spike train as
\begin{eqnarray}
\bra\rho_s(t)\ket ~&=&~ \sum_m ~\langle e^{i 2\pi m 
\Phi(t)
}
\dot{\Phi}(t) \rangle. \label{rav}
\end{eqnarray}
For a stationary noise $n(t)$, and time long enough for the system to be
in a steady state, this average is independent of the time $t$. 
We assume that the noise has a short correlation time $\tau_n$; this implies
that the noise can have an arbitrary distribution at each time,
but that it is
uncorrelated with the noise value at a time much later.
For $t$ larger than the noise correlation time, $\tau_n$,
$\Phi(t)$ is an integral of many independent random variables, and
is approximately Gaussian by the central limit theorem.
Then, one can substitute the average of the exponent by the 
exponent of the two first cumulants:
\be
\bra e^{i2\pi m \Phi(t)}\ket 
\approx e^{i2\pi m\bra \Phi(t)\ket - 2 \pi^2 m^2\bra \delta \Phi(t)^2\ket}
\label{Gauss_approx}
\ee
\noindent where $\langle\delta \Phi(t)^2\rangle=
\langle \Phi(t)^2\rangle - \langle \Phi(t)\rangle^2$.
The first two cumulants of $\Phi(t)$ are:
\begin{eqnarray}
\bra \Phi(t)\ket ~&=&~ \int_0^t\langle f[s\!+\!n(t')]dt' \rangle
=rt \label{cumulants} \\
\bra \delta \Phi(t)^2 \ket &=& \int_0^t\int_0^t dt' dt''
\langle \delta f[s\!+\!n(t')] \delta f[s\!+\!n(t'')] \rangle.
\nonumber
\end{eqnarray}
But $f$ is correlated only over a short time, on the order of $\tau_n$
and therefore the double integral can be approximated by
\begin{eqnarray}
\langle \delta\Phi(t)^2\rangle &\approx&
\int_0^t \!d\bar{t}\int_{-\infty}^{\infty} \!d\zeta
\langle \delta f(s+n(\bar{t}\!+\!{\zeta\over 2}))\delta 
f(s+n(\bar{t}\!-\!{\zeta\over2}))\rangle\nonumber\\
&\approx& Dt,
\end{eqnarray}
with $D=\tau_n \langle\delta f^2\rangle$.
Using Eq. (\ref{Gauss_approx}) in (\ref{rav}), and taking the time $t$
arbitrarily large, we find 
\be
\bra\rho(s)\ket ~=~ \bra f(s+n)\ket ~=~ [f*G] (s),  \label{conv}
\ee
\noindent 
where $*$ denotes convolution and $G$ is the distribution function of
the noise at a given point of time. 
For a stationary noise distribution, this
implies that the average firing rate as a function of signal $s$
has the form of $f(s)$ smeared by the noise. 
This result is independent of the exact form of the function $f$ or of
the noise distribution. It relates the $f/I$ curve to a measurable
quantity, an average firing rate in the presence of noise; 
in the measured quantity, much of the
fine details of $f$ will be smeared by the noise. 

In order to compare to data,
we must now identify the stimulus $s(t)$ in the
experiment. If the neuron is isolated in a dish and current is injected
directly into the cell, 
then $s(t)$ should be naturally identified with this current.
In experiments with an intact sensory system
one would like to connect
$s(t)$ with the external stimulus. In the following discussion we
will use as our test case the visual system of the
blowfly. 
In our experiment, a live immobilized fly views various visual stimuli,
chosen
to excite the response of the cell H1.  This large neuron is located
several
layers back from the eyes, and receives input through connections to
many other cells.
It is identified as a motion detector, responding optimally to
wide-field rigid horizontal motion, with strong direction selectivity
\cite{Facets89a,Facets89b}.
The fly watches a
screen with a random pattern of
vertical dark and light bars, moving horizontally with a velocity $s(t)$.
We record the electric signal of H1 extracellularly,
and register
a sequence of spike timings $\{t_k\}$ \cite{Ruyter95}.

An advantage of this system 
is that we have empirical knowledge of what stimulus feature is
relevant to this cell: it responds to wide field motion in the
horizontal direction. 
Thus we may identify the input to the
cell directly with a one dimensional
external signal, the motion of the pattern
on the screen. 
Figure 1 shows the firing rate averaged over many presentations of the
same stimulus, both as a function of time (Fig. 1a) and as a function
of the instantaneous value of $s(t)$ (Fig. 1b).  
If the velocity on the screen is slowly varying in time, 
the firing rate of H1 follows this velocity closely:
Fig. 1a shows the time-dependent firing rate
$r(t)=\bra\rho[s(t)]\ket$ of H1 in response to a random signal.
Since the signal varies very slowly,
one may use Eq. (\ref{conv}) locally, and if we plot 
the firing rate as a function of the instantaneous value
of $s$, we see the smoothed response function 
$f(s)$ (Fig. 1b).
If the signal varies rapidly, 
filtering mechanisms at early stages of the visual pathway become
important, and $s(t)$ which directly drives the cell is a modified 
version of the velocity on the screen. 
In this case, it is more difficult to map out directly the response
curve \cite{adapt}. It should be noted, however, that we present the response
function in Fig. 1 only for purpose of illustration; in what follows we
shall assume that this function is well defined, but we will not need to
know its exact shape. Moreover, our results will show that to account
for the statistical properties of the spike trains very little
information about the response function is required. 

{\narrowtext
\begin{figure}[h]
\begin{center}
\epsfxsize=9cm
\epsfbox{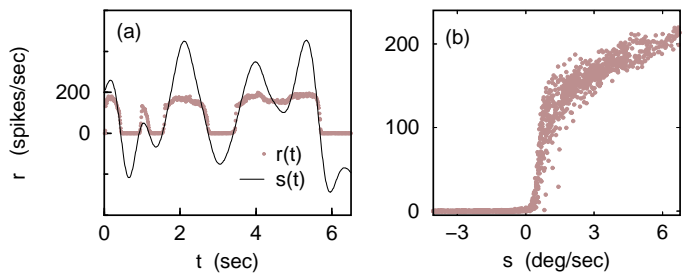}
\end{center}
\caption{
Firing rate of H1 as a function of time, averaged over trials:
$r(t)=\langle\rho(t)\rangle$ [dots],
compared to the input
signal $s(t)$ [solid line], for a slow signal (a) and a fast signal
(c).
We repeat the signal many times to obtain a sampling of the noise
ensemble.
The units of velocity are
spacings of the compound eye's lattice (ommatidia) per second.
}
\label{fig:1}
\end{figure}
}

It should be noted that the response function of the neuron
$f(\cdot)$ is not a fixed property, but may change with
external conditions. The theory presented here is valid within a
steady state, in which $f(\cdot)$ takes a particular form and does not
change with time. Rather than
being a limitation, the context-dependence of $f(\cdot)$
opens the possibility to investigate
adaptive changes in the neural response, when adjusting to different
steady states \cite{adapt}.

\section{The universal regime}

In this section we present the statistical theory for the case of a
constant input signal, $s(t)\!=\!s$, and a random short correlated
noise of arbitrary distribution.  The result is a renewal process with
special symmetry properties, reflecting the underlying neuronal
oscillator.  We provide explicit formulas for the correlation function
and the number variance.  
As will be shown in later sections, the results obtained
here are valid in a limited time range if the
stimulus $s(t)$ is time dependent.

\subsection{Distribution of Intervals}

The distribution of inter--spike intervals is the 
distribution of times for which $\Phi(t)=1$. Due to the unidirectionality
of the phase diffusion, these times are unique, and so the probability
density is simply 
\begin{eqnarray}
P(t)~=~ \Bigg\langle
 \delta(t-\Phi^{-1}(1))\Bigg\rangle ~&=&~ \Bigg\langle
\delta(\Phi(t)-1) {\dot \Phi}(t))\Bigg\rangle. \label{non-neg}
\end{eqnarray}
It is more convenient to calculate the cumulative
distribution, 
\be
F(t)~=~\int_0^t P(t')dt' 
\ee
which can be expressed in terms of the step function, $\Theta$,
and its Fourier transform:
\be 
F(t)=\bra \Theta(\Phi(t)-1)\ket~=~\int_{-\infty}^{\infty} \frac{dp}{2\pi i
p} e^{-i2\pi p}\bra
e^{i2\pi p \Phi(t)}\ket. \label{F(t)}
\ee
\noindent Using the Gaussian approximation 
(\ref{Gauss_approx}) for $\Phi(t)$, 
we find that
\begin{eqnarray}
F(t)~&=&~
\int_{-\infty}^{\infty}\frac{dp}{2\pi ip} e^{i2\pi pr t-2\pi^2 p^2 Dt}
\nonumber \\
&=& 
\frac{1}{2}\left( 1+\E\left(\frac{rt-1}{\sqrt{2Dt}}\right)\right)
\label{cum}
\end{eqnarray} 
and the interval density is
\be
P(t)=\frac{rt+1}{\sqrt{8\pi D t^3}} ~e^{-(rt-1)^2/2Dt}. \label{pdim}
\ee
\noindent
In Appendix 1 this result is derived for a discrete sum of non-negative
random variables, directly from the central limit theorem.
The density (\ref{pdim})
depends on two parameters, the average firing rate $r$ and
the diffusion coefficient $D$, both 
of dimensionality [time]$^{-1}$. In the derivation,
we used only the non--negativity of the frequency integrator to write
down Eq. (\ref{non-neg}), and the Gaussian approximation 
(\ref{Gauss_approx})
for $\Phi(t)$; therefore 
changes in the model which retain these properties will not affect
the interval density. 
Eq. (\ref{pdim}) is similar to the first passage time 
of the Wiener
process \cite{Tuckwell88,Gerstein64}: 
it has the same exponent, but this exponent is multiplied by a
different function of $t$. As will be shown below, this results in
significantly different symmetry properties of the function.

To understand the qualitative properties of the point process, 
it is convenient to examine it in dimensionless time
units, namely to define the time such that the average rate is 1.
We denote this time as $x\!=\!rt$.
In this representation, the statistics depend on
one dimensionless parameter, $\gamma$, 
\be
\gamma = \frac{D}{r} = \frac{\tau_n \bra\delta f^2\ket}{\bra f\ket}
\label{gamma}
\ee
and the interval density is 
\begin{eqnarray}
P(x) = \frac{x+1}{\sqrt{8\pi \gamma  x^3}}
e^{-(x-1)^2/2\gamma x}. \label{ps}
\end{eqnarray}
The parameter $\gamma$ governs the decay of the interval
density both near the origin and at large $t$. For small $\gamma$ these
decays are strong, indicating a narrow density, and for large $\gamma$ 
they are weaker and the distribution is broader.
More formally, 
the moment generating function,
$G(\lambda)=\bra e^{-\lambda x}\ket$, 
of (\ref{ps})
can be calculated using the 
integral representation,
\be
P(x) = \int_{-\infty}^{\infty} dp (1+i\pi\gamma p)e^{i2\pi(x-1)p
-2\pi^2 \gamma p^2 x}.
\ee
The result is
\be
G(\lambda)=
\frac{1}{2}\left(1+\frac{1}{\sqrt{1+2\lambda\gamma}}\right)
e^{\frac{1}{\gamma}(1-\sqrt{1+2\lambda\gamma})},
\ee
and the first two moments are
\begin{eqnarray}
\bra x\ket &=& 1+\gamma/2\\
\bra\delta x^2\ket &=& \gamma+
\frac{5}{4}\gamma^2.
\end{eqnarray}
Note that the average interval length is not equal to the inverse of
the average rate, which is 1 in our units. In general, inversion does
not commute with averaging; for small $\gamma$, however, the inverse
average and the average of the inverse are similar.

Defining the coefficient of variation by 
$\sqrt{\bra\delta x^2\ket}/\bra x\ket$
\cite{Hagiwara54}, it is seen that the
coefficient of variation is approximately proportional to $\sqrt \gamma$
for small $\gamma$, and to $\gamma$ for large $\gamma$. It
can take on
arbitrarily small and large values dependent upon $\gamma$.
This is a more quantitative way of seeing that
our family of distributions interpolates between
low--variability (or regular) and high--variability (or irregular)
limits.
It should be noted that this distribution arises from a simple
integration model of many independent inputs \cite{Softky93}.
As noted already by several authors \cite{Troyer97,Gutkin98},
it is not only the properties of
the inputs but also of the internal neural response that determine the
degree of irregularity of a spike train. In our model, 
the parameter controlling this irregularity (\ref{gamma}),
accounts for both these effects.

The fact that $\gamma$ controls the behavior of the density 
(\ref{ps}) at both tails, takes the quantitative form of an
invariance under the 
transformation $x\to 1/x$, with the Jacobian properly accounted
for. This implies that the cumulative distribution of the intervals
between successive spikes is identical to the distribution of inverse
intervals, when both are measured in dimensionless units.
Since our spike train is a renewal process, this invariance of the
interval distribution implies an invariance of the process as a whole.
In general, for a
point process with time intervals $\{x_i\}$, one
may construct the dual process with intervals $\{1/x_i\}$; this has
the natural interpretation of local frequencies. 
The process defined by (\ref{ps}) is {\em self-dual}: all statistical
properties of the dual process are identical to the original one. 
For the process defined by (\ref{pdim}), this invariance holds 
up to a global rescaling
of the axis.
Fig. 2 shows that this is indeed a property of the measured data: the
cumulative distributions of intervals and inverse intervals overlap when
plotted in dimensionless units. 

{\narrowtext
\begin{figure}[h]
\begin{center}
\epsfxsize=8cm
\epsfbox{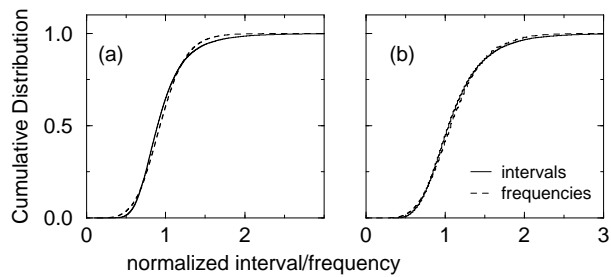}
\end{center}
\caption{
A symmetry of the spike train point process: the distribution of
intervals
between neighboring spikes is the same as the distribution of the
inverse intervals (local frequencies),
up to a constant scaling factor. Data are shown
from two experiments with constant velocity stimuli, of magnitude 10.5
(a) and 0.16 (b) deg/s.
}
\label{fig:2}
\end{figure}
}

A generalization of the interval distribution is the {\em scaled interval
distribution of order $k$}, defined as the distribution of
intervals between pairs of spikes that have exactly $k-1$ 
other spikes in between them.  In their landmark paper, Gerstein and
Mandelbrot  (1964)
observed that the scaled interval
distributions of low orders in spike trains from the cat cochlear
nucleus, have a similar shape. This observation
motivated them to suggest the random walk model for the membrane
voltage.
In our model, we can calculate the scaled interval
distribution directly: it is the distribution of times for which $\Phi(t)=k$,

\begin{equation}
F_k(t) ~=~ \frac{1}{2}\left(
1+\E\left(\frac{rt-k}{\sqrt{2Dt}}\right)\right). \label{cumk}
\end{equation}

\noindent In this notation, 
the interval distribution of Eq. (\ref{pdim}) is the scaled
distribution of order $1$.  Figure 3a shows the first three scaled
interval distributions, as measured experimentally, together with Eq.
(\ref{cumk}). The two fitting parameters of the theory, the
average rate $r$ and the diffusion constant $D$, are fitted once for all
three graphs.  According to the observation of Gerstein and Mandelbrot,
these curves should have the same shape after rescaling the time
axis to dimensionless units $rt/k$.
Figure 3b shows the scaled interval distributions in dimensionless time
units. These curves have a similar shape, but do not quite overlap. As is
easily seen from Eq. (\ref{cumk}), the transformation $t\to t/k$ gives a
function of the same general form, but with a different value of $D$.
Thus the family of curves $F_k(t)$ with parameters $r,D$ obey the
following equation:
\be
F_k^{(r,D)}(t/k)~=~F_1^{(r,\frac{D}{k})}(t), \label{F_relation}
\ee
where the superscripts denote the dependence on the parameters.
In  Figure 3b, it is clearly seen that the steepness of the curve
increases with increasing $k$, consistent with a decrease in the
diffusion constant $D$.
The point $rt/k=1$ is
where the distributions cross, in agreement with the theoretical equation
(\ref{F_relation}).

{\narrowtext
\begin{figure}[h]
\begin{center}
\epsfxsize=8cm
\epsfbox{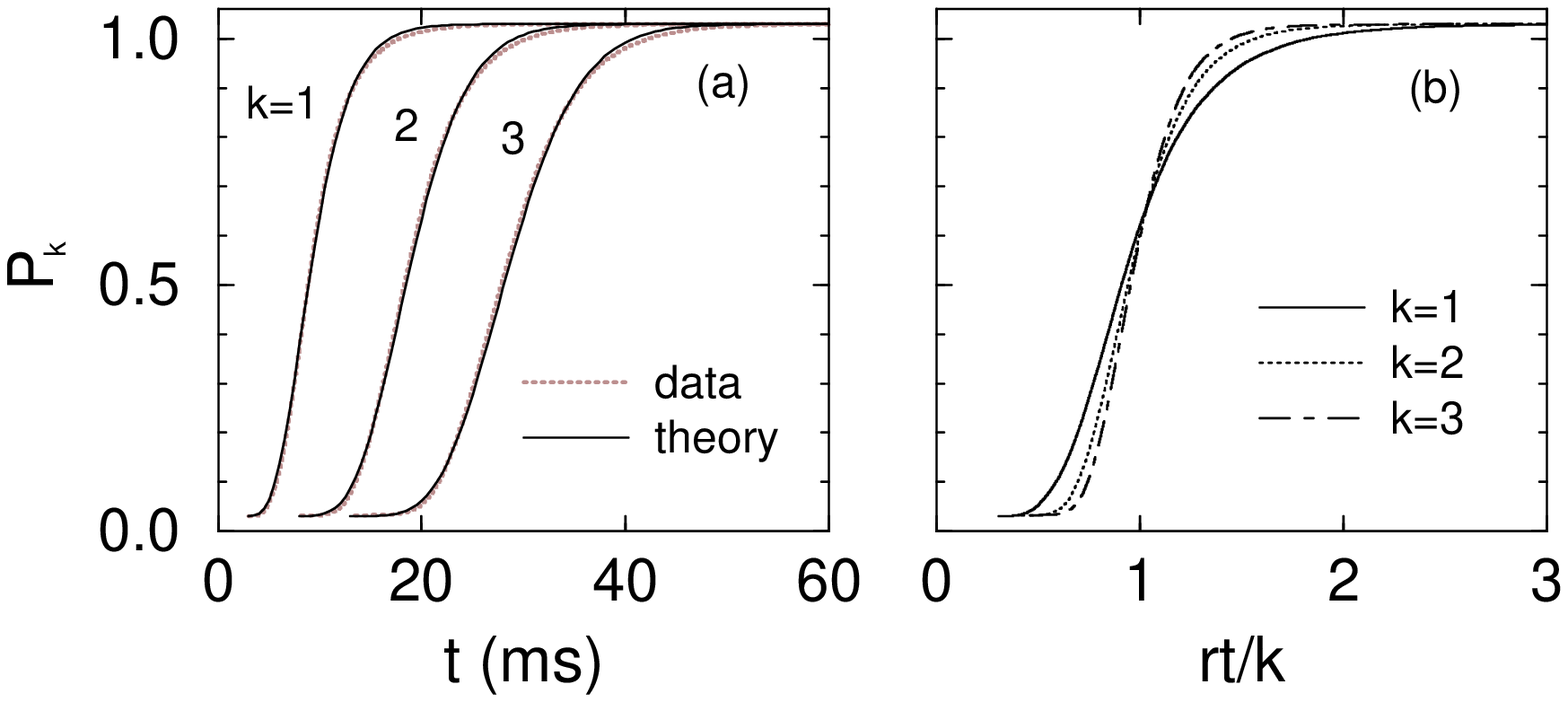}
\end{center}
\caption{
(a) Scaled interval distributions of orders $k=1,2,3$, representing
the probability of finding
a pair of spikes at a given time interval, with exactly $k-1$ spikes
in between. Data are shown in grey dotted lines, while theory (Eq.
\ref{cumk}) is shown in a solid black line. Fitting parameters of the
theory, $r$ and $D$, are fitted once for all three curves.
(b) Scaled interval distributions in dimensionless time units. In
agreement with Eq. (\ref{F_relation}), these curves are similar but
have
slightly different slopes, corresponding to an effective value of $D$
which depends on $k$. The curves cross at $rt/k=1$, as predicted by
Eq. (\ref{F_relation}).
}
\label{fig:3}
\end{figure}
}

The derivation of the interval density is independent of many
microscopic details of the neuron and of the noise, and it is therefore
expected to
describe correctly the behavior of many different systems. 
Figure 4 shows a comparison of the interval distribution (\ref{pdim})
with experimental
data from different parts of the visual system in several animals.
Figure 4a shows data measured in our experiment on the motion sensitive
neuron H1 in the visual system of the blowfly. In this experiment, the
fly watched a random pattern of dark and light bars moving at a constant
velocity of $\sim 0.16$ deg/sec. The best fit to the data was found with
an irregularity parameter of $\gamma=0.1$.  Figure 4b is adapted from
data published by Robson and Troy 
(1987). In this experiment, a
stationary sinusoidal grating was presented to an anesthetized cat, and
spikes were recorded from neurons in the retina. The
particular neuron these data were recorded from was identified as a
``Q-type'' neuron, characterized by regular spiking. The best fit of Eq.
(\ref{pdim}) was obtained with $\gamma=0.015$. 
Figure 4c shows data
measured from isolated goldfish retina by Levine and Shefner (1977)
in darkness. These data are well described by Eq.
(\ref{pdim}) with $\gamma=0.1$.
Figure 4d contains data
measured by Cattaneo et al. (1981) from visual cortex of
anesthetized cat, in response to a drifting sinusoidal grating.  This
interval density is best fit with $\gamma=0.3$. 

{\narrowtext
\begin{figure}[h]
\begin{center}
\epsfxsize=8cm
\epsfbox{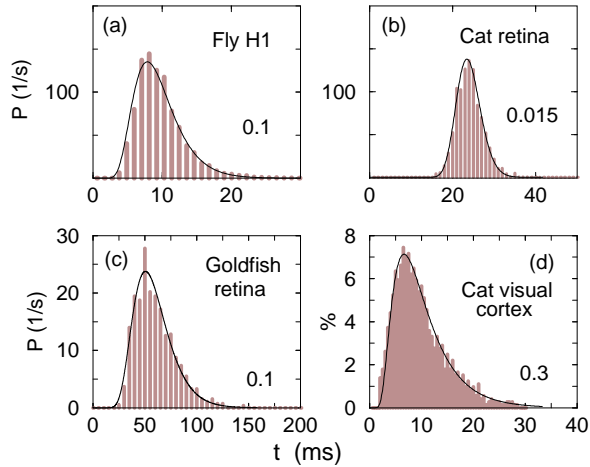}
\end{center}
\caption{
Distribution of intervals between neighboring spikes, experiments and
theory.
The data are from different sensory systems in different animals:
(a) fly visual system (lobula plate motion detector), (b) Cat retina,
(c) goldfish retina
(d) cat visual cortex.
For more details about these published data see text.
The y-axis is the probability of seeing an interval, in units of 1/s
(a-c), and the relative probability in percent (d).
The solid black lines are a fit of Eq. (\ref{pdim}),
and number in each panel indicates the value of the irregularity
parameter
$\gamma$ in the fit.
}
\label{fig:4}
\end{figure}
}

These data indicate that the classification of spike trains 
to {\em universality classes} according to
the irregularity parameter $\gamma$ is a useful one, and can be applied
to many systems. Systems with very different microscopic properties can
belong to the same universality class, 
and their interval density is well described by a theory with one
dimensionless parameter. It will be shown in section 4, that this
classification can be applied also under conditions of rich dynamic
stimuli; in this case, the internal irregularity parameter $\gamma$
describes the statistical properties on short time 
scales.

\subsection{Correlation function}

An important statistical property of the spike train is its
(auto)-correlation function, $\langle \rho(t)\rho(0)\rangle$. Whereas many
models have been used to calculate the interval density, less attention has
been paid to
the correlation function \cite{Gestri75}. In Appendix B 
we derive the correlation function 
under the assumption that the noise correlation time $\tau_n$ is much shorter
than the typical interval between spikes $1/r$. The result is:
\begin{eqnarray}
\langle \rho(t)\rho(0)\rangle 
& = & r\delta(t)~+~
r \sum_{k\neq 0} P_k^{(r,D)}(|t|)
\label{Rdim}
\end{eqnarray}
where $P_k^{(r,D)}(t)$ are the densities derived from the scaled interval
distributions of Eq. (\ref{cumk}),
\be
P_k^{(r,D)}(t) ~=~ \frac{d}{dt} F_k^{(r,D)}(t)~=~ \frac{rt+k}{\sqrt{8\pi D
t^3}}~e^{(rt-k)^2/2Dt}. 
\ee
It is convenient to think about the probability per unit time
of finding a spike at
time $t\neq 0$, conditional on the event that a spike is found at time $t=0$.
This quantity $R(t)$ is proportional to the
correlation function in the region $t\neq 0$,
and is called the conditional rate:
\begin{eqnarray}
R(t) ~&=&~ 
\sum_{k\neq 0} P_k^{(r,D)} (|t|) ~\equiv~ R_U^{(r,D)}(t). \label {RU}
\end{eqnarray}
The density labeled $k$ is the
probability per unit time for finding a pair of
spikes separated by a 
time $t$ with exactly $(k-1)$ spikes in between. 
These independent events, when added together, give the total 
probability per unit time to find a pair of spikes separated by
a time $t$, which is just the conditional rate.
For small $k$ these individual densities $P_k^{(r,D)}$
are narrow and their peaks 
can be resolved, and as $k\to\infty$ they smear and overlap to give
asymptotically $R(t)\sim r$. 
The degree of regularity, $1/\gamma$, is associated with the number of
densities which can be resolved.
The notation $R_U^{(r,D)}(t)$ is introduced
to emphasize that this is a universal function, independent of the
detailed neuronal response $f(s)$, and of the detailed properties of the
noise.
Figure 5 shows the conditional rates for experiments with
constant velocity, together with the best fit to Eq. (\ref{Rdim}). 
The constant value of the velocity stimulus $s$ increased among
parts (a-d) of the figure; the average firing rate $r$ increases and the
irregularity $\gamma$ decreases. 
This is clearly due to refractory
effects: as the firing rate increases, the repulsive interaction among
the spikes becomes more important, and the spike train becomes more
stiff, causing a more regular behavior. The form of the repulsive
interaction and the function describing the correlation function,
however, remain the same. 

{\narrowtext
\begin{figure}[h]
\begin{center}
\epsfxsize=8cm
\epsfbox{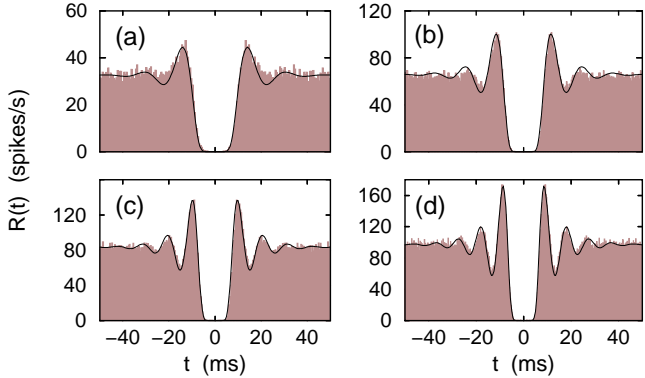}
\end{center}
\caption{
Conditional rate for the spike train in experiments with a constant
velocity stimulus. Data are shown by a gray histogram, calculated
directly from the spike times by binning them into 4 ms bins.
Theoretical expression
(Eq. \ref{Rdim}) is shown by a solid black line. The four parts of the
figure
correspond to different values of the constant velocity stimulus: 0.7
(a),
2.6 (b), 10.5 (c) and 42.2 (d) deg/sec. As the
motion signal becomes stronger, the average firing rate increases and
the irregularity parameter decreases (see also Fig. 6).
}
\label{fig:5}
\end{figure}
}

\subsection{Number variance}

The number variable $N(t)$ is a random function which 
counts the number of spikes in the time window $[0:t]$.
Defined by
\be
N(t)~=~ \int_0^t \rho(t')dt', \label{Nt} 
\ee
it provides a useful, less detailed, characterization of the 
spike train point process.
To study the statistics of this
variable, we write it as:
\be
N(t)=\sum_{k=1}^{\infty}\Theta(\Phi(t)-k)=\mbox{Int}(\Phi(t))
\ee
where Int$(y)$ is the integer part of $y$. 
Its Fourier representation is
\be
N(t)=\Phi(t)+\frac{1}{2}-\varphi +\sum_{m\neq 0}\frac{1}{2\pi i m}~e^{i2\pi
m(\Phi(t)+1-\varphi)} \label{num_four}
\ee
where $\varphi$ is uniformly distributed on $[0,1]$,
accounting for a random position of the first spike in a window (see
App. C).
The number mean is  $\langle N(t)\rangle = rt\nonumber$, and
around this mean $N(t)$ fluctuates.
Since the correlation function of spikes is of finite range,
the long time asymptotic behavior of the number variance
is diffusive: $\sigma^2_N=\langle \delta N^2(t)\rangle\sim Dt$
\cite{Spikes97}. 
It is of interest, however, to
derive a complete expression for this quantity also for short times, 
where correlations between spikes are important.
Using (\ref{num_four}), it is shown in Appendix C that 
\begin{eqnarray}
\sigma^2_N &=& Dt +\sum_{m=1}^{\infty}\frac{1}{(\pi m)^2} \left[1-\cos(2
\pi mrt)~e^{-2\pi^2 m^2 Dt}\right]. \label{NV}
\end{eqnarray}
Written as function of the number
mean, i.e. as a function of the dimensionless time variable
$x\!=\!rt$, the number variance is:
\be
\sigma^2_N ~=~  \gamma x +
\sum_{m=1}^{\infty}\frac{1}{(\pi m)^2}\left[1-\cos(2\pi m x)e^{-2\pi^2
m^2 \gamma x}\right]. \label{num_var}
\ee
Figure 6 shows the number variance as a function of the number mean,
as calculated from an experiments with constant velocity stimuli. 
Part (a) shows a scatter-plot of the values obtained in various windows
in the experiments, with a line showing the average. Part (b) of the
figure shows a detailed view of one such average plot, with the
theoretical prediction 
Eq. (\ref{num_var}) in solid
black line. The number variance of a Poisson process is 
shown for comparison.
The ``diffusion constant'' in the number variable is
$\gamma$, implying again its role as an irregularity parameter: the
more stochastic the point process, the faster is the diffusion in the
number variance. Similar to Figure 4, a higher constant signal induces
a higher firing rate and a lower irregularity of the spike train.
Although the data
presented here have $\gamma<1$, this is not a fundamental
property of the theory, and in general, $\gamma$ can take on also values
larger than 1, resulting in a 
``super-Poisson" behavior.

{\narrowtext
\begin{figure}[h]
\begin{center}
\epsfxsize=8cm
\epsfbox{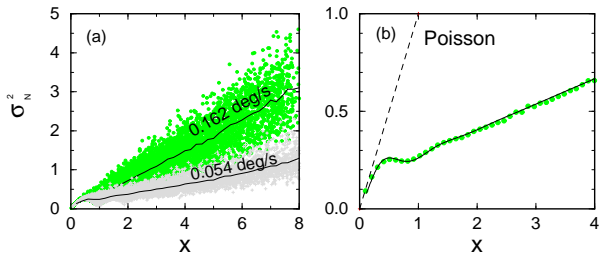}
\end{center}
\caption{
Number variance as a function of number mean. (a) Scatter plots of
variances, as calculated in different time windows in the experiment.
Data are shown from two experiments, with constant velocities of 0.054
and 0.162 deg/s.  Solid lines indicate the average value of the
variance
for a given mean. (b) Comparison to theory: the average value at each
number mean, is compared to the theoretic formula
(Eq. \ref{num_var}), which is shown by a solid black line.  In this
experiment, the velocity was 0.162 deg/s.
}
\label{fig:6}
\end{figure}
}

\subsection{Irregularity and stiffness of spike trains}
We have presented a statistical theory for a frequency
integrator model under constant stimulation and short-range correlated noise.
This results in a renewal point process, with the density of intervals given by
Eq. (\ref{pdim}).
This process is characterized by an irregularity parameter
$\gamma$, which depends both on the variance of the noise and on the
sensitivity of the frequency response to noise; it is the depth of frequency
modulation, resulting from these two effects, that determines the irregularity
of the process. This one-parameter family of processes can describe many
different spike trains, ranging from almost periodic to almost Poisson.

In the data presented above, a correlation was found between the global
average firing rate and the irregularity $\gamma$. Figure 7 shows a
plot of the different values of $\gamma$ obtained by fitting to the
equations in this section, for two sets of experiments. The
irregularity decreases roughly linearly with the firing rate, with a
saturation at very high firing rate. This is the limit where
the absolute refractory period is approached. The simple relation between
$\gamma$ and $r$ holds
only for the case of constant stimuli, where frequency modulations are
essentially induced by noise. When these modulations are affected also by
a time varying sensory stimulus, very different behaviors is found
(see next section).

{\narrowtext
\begin{figure}[h]
\vspace{0.2cm}
\begin{center}
\epsfxsize=7cm
\epsfysize=5.cm
\vspace*{-0.2cm}
\epsfbox{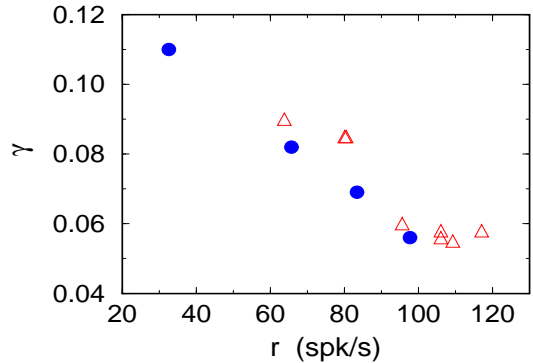}
\end{center}
\caption{
Dimensionless irregularity parameter, $\gamma$,
as a function of the average firing
rate $r$. The different symbols correspond to two different
experiments, performed each with a different set of constant velocity
signals.
}
\label{fig:7}
\end{figure}
}

\section{Time dependent stimulus}
In the previous section, we considered the case of a constant input
signal, $s(t)\!=\!s$. More generally signals coming into the system have a
temporal structure, and additional time scales enter the problem.
If $s(t)$ is slowly varying relative to $1/r$,
universal behavior is expected
on short times, and slower
modulations will appear on longer times. In this section we consider
the statistical properties in the case of a slowly varying input
signal. We show that the conditional rate can be approximated by a product
of the universal function Eq. (\ref{RU}), 
and a slowly varying envelope reflecting the
temporal correlations of the input signal.
This envelope
is calculated for some simple cases. It is shown that the theory fits
the data very well, even when
the time scale separation required in theory is
only marginally satisfied by the experimental conditions.

\subsection{The Telegraph Approximation}

Consider a random input signal $s(t)$,
that can take on positive as well as negative values.
Figure 8 shows a segment of
some spike trains recorded from H1 
in response to such a stimulus, which was repeated
many times. The most striking effect in the figure 
is the existence of wide regions with spikes,
and wide
regions which are empty; the typical time for these regions is
much larger than the interspike time.
This partition into regions is a consequence of the direction selectivity
of the H1 neuron: it fires when the effective stimulus is in its preferred
direction, and is inhibited by stimulus in the opposite direction.
Although the velocity stimulus one the screen 
varies rapidly in this experiment (every 2ms an independent value is
chosen), the spiking and quiet
regions in the spike trains
have a much slower typical time scale.  This results from intermediate
processing: the velocity on the screen is not identical to the
effective stimulus driving the cell, since it is
filtered by the photoreceptors and other elements in
the visual pathway.  

To describe the phenomenon of spiking regions and quiet regions,
we imagine the spike train 
to be multiplied by a telegraph signal, which keeps track of
the algebraic sign of the effective stimulus:
\begin{equation}
\rho(t)\approx \rho^{S}(t) \rho^{E}(t), \label{Product}
\end{equation}
where $\rho^{E}(t)=
\Theta[s(t)]$ is the telegraphic envelope of the spike train.
Figure 8 shows an illustration of the telegraph signal, which demonstrates
that the the time scale of the effective stimulus sign change is longer
than the typical spike time, $1/r$. 

{\narrowtext
\begin{figure}[h]
\vspace{0.2cm}
\begin{center}
\epsfxsize=7cm
\epsfysize=5.cm
\vspace*{-0.2cm}
\epsfbox{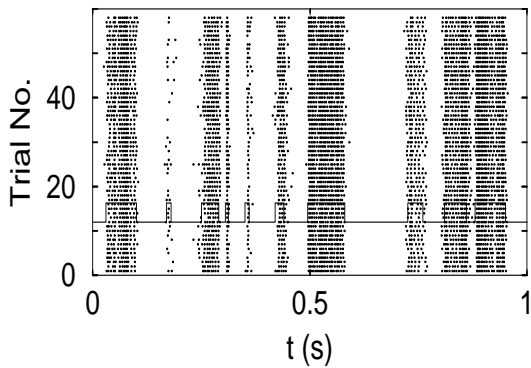}
\end{center}
\caption{
Response of the H1 neuron to repeated presentations of a
time-dependent
stimulus. Each line shows the response to one of the presentations,
with
the origin $t=0$ corresponding to the onset of the stimulus; each dot
represents a spike . The
black solid line is an illustration of
the telegraphic envelope that multiplies the
spike train, due to the strong direction selectivity of the neuron.
}
\label{fig:8}
\end{figure}
}

Assuming that the telegraph envelope is statistically independent of the
short time structure in the spike train, 
one may write the correlation function as a product,
\begin{eqnarray}
\langle \rho(t)\rho(0)\rangle &=& \langle \rho^{S}(t)\rho^{S}(0)
\rho^{E}(t)\rho^{E}(0)\rangle\nonumber\\ &\approx& \langle 
\rho^{S}(t)\rho^{S}(0)
\rangle \langle \rho^{E}(t)\rho^{E}(0)\rangle. \label{stat_ind}
\end{eqnarray}
Now brackets $\langle\cdots\rangle$ denote
averaging over both the noise and the random stimulus. To perform the averages,
we use the separation of time scales: imagine dividing the
time axis into blocks of size $\tau_s$, the correlation time of the
effective
stimulus. 
Performing first the
average over the noise in each block separately, 
the first term in the product gives the
correlation function of Eq. (\ref{Rdim}), with the parameters $r,D$
determined by the local value of $s$: 
\begin{eqnarray}
\langle\rho^S(t)\rho^S(0)\rangle \approx \Bigg\{ \begin{array}{l}
r(s)~R_U^{(r(s),D(s))}(t),~~~~~~~~~~\hfill s>0\\
              0,\hfill s\le 0
\end{array}
\end{eqnarray}
for $t\neq 0$.
If the response is saturated, $r(s) \approx r \Theta(s)$, and
the firing rate
does not change much
inside each spiking region; 
moreover the diffusion constant $D$ is determined mainly
by properties the noise, and is the same in all the spiking
regions. Therefore, on averaging the first term over $s$, one has 
$\alpha R_U^{(r,D)}$,  where $\alpha$ is the coverage fraction,
defined as the fraction of time in
which the stimulus is positive ($\langle\rho^E(t)\rangle=\alpha$).

The second term in the product (\ref{stat_ind})
is a correlation function of the input
signal as seen through the telegraphic envelope:
\begin{eqnarray}
\langle\rho^E(t)\rho^E(0)\rangle =  \langle
\Theta[s(t)]\Theta[s(0)]\rangle.  \label{sign_corr}
\end{eqnarray}
\noindent 
Special care should be taken around the point $t=0$, since it is
affected by the delta function singularities. The 
whole correlation function then takes the form
\begin{eqnarray}
\langle\rho(t)\rho(0)\rangle~~ \approx~~ 
r \alpha \delta (t) ~+~ r R_U^{(r,D)}(t)
~\langle\Theta[s(t)]\Theta[s(0)]\rangle,    \label{product1}
\end{eqnarray}
and  the conditional rate is
\be
R(t) ~~=~~ 
R_U^{(r,D)}~\langle\Theta[s(t)]\Theta[s(0)]\rangle.
\label{product2}
\ee
This formula expresses the conditional rate 
as a product of two terms.  
The first term reflects internal properties of the noise and of the
neuron, similar to the result of the previous section; 
$R_U$  is
parameterized by an effective rate $r$ and diffusion constant 
$D$. This function
has an oscillatory structure on a time scale $1/D$,
which thus defines the {\em universal regime} of the correlation function,
in which properties of the inputs do not have an important effect.
The second term contains information about statistics of the incoming 
stimulus, as seen through the nonlinear response of the neuron. It modulates
the universal function with a slower structure. Intuitively, the condition of 
time scale separation can be understood as follows: on short times, $t\leq
1/D$, the envelope is almost constant and the oscillations of the universal
part are visible. As these oscillations decay, on times $t>1/D$, the stimulus
induced structure sets in. The independence between the two factors
affecting the probability of finding a spike at time $t$ given a spike
at time $0$, results in a product form.

Eq. (\ref{sign_corr}) indicates that in the telegraph approximation, the 
envelope of the correlation function depends on
the statistics of the zero-crossings of the stimulus. 
These statistics for a random continuous function are
in general very
difficult to calculate \cite{Rice54}. Here we need only the correlation
function of the algebraic sign of a random signal, and we can proceed in two
ways. In the next sub-section we consider the case
in which zero crossings are independent, and the resulting correlation
function is a simple exponential. 
Another simplification occurs if the incoming 
stimulus has Gaussian statistics,
and in this case one may derive an exact formula 
for the correlation function
of the nonlinear rate.

\subsection{Random independent spiking regions}
Let us first assume that the spiking regions occupy random independent 
positions along the time axis, with 
lengths $\Delta$ drawn independently from some distribution.  
This is a crude approximation that may not be justified for many
experimental conditions.  However, it is the simplest type of telegraphic
signal, and is characterized by a small number of parameters; therefore
we use this approximation as a starting point. 
The correlation function of
such a telegraph signal is derived in Appendix D, and the result is,
\begin{eqnarray}
\langle \rho^E(t)\rho^E(t')\rangle &=& 
\alpha^2 
\frac{\alpha}{\mu}
\int_{|t\!-\!t'|}^{\infty} p(\Delta) 
(\Delta-|t\!-\!t'|)d\Delta \label{env_corr}
\end{eqnarray}
\noindent where $p(\Delta)$ is the distribution of lengths of positive
regions in the telegraph signal, and $\mu$ is the average length of such
a positive region.
The second term clearly decays for
$(t\!-\!t')\!\to\!\infty$ for a well behaved $p(\Delta)$, 
and asymptotically there remains only the
square of the average signal, $\langle\rho^E\rangle^2\!=\!\alpha^2$.
For the case of an exponential distribution of positive regions,
\begin{eqnarray}
p(\Delta)= \frac{1}{\mu}e^{-\Delta/\mu},
\end{eqnarray}
\noindent
the correlation function is also exponential, 
\begin{eqnarray} 
\langle \rho^E(t)\rho^E(t')\rangle =
\alpha^2 +\alpha \mu e^{-(t\!-\!t')/\mu}. \label{exp_tlgrph}
\end{eqnarray}
\noindent
The exponential decay of the envelope, Eq. (\ref{exp_tlgrph}), gives a
good fit to many of the measured data sets. It has additional fitting
parameters,  $\alpha$ and $\mu$, which are the coverage fraction and
average length of the positive regions in the telegraph envelope of the
spike train.
It uses no prior knowledge of the input stimulus, and relies primarily on the
direction selectivity of the response.
Figure 9 shows the
correlation function calculated from spike trains, in two experiments with
random signals, of 20Hz  (a,b) and 500Hz (c,d) bandwidth.
Although in these experiment we know the
properties of the visual stimulus presented to the fly, this
stimulus is rapid, and pre-processing takes place in earlier stages of
the visual pathway; the effective signal entering the H1 neuron is 
therefore a filtered and probably distorted version of the motion on the
screen.  Therefore, we try the simple picture of the telegraph 
approximation, rather than
rely on the detailed properties of the visual stimulus. 
The black solid line in Figure 9 is a fit to a product of 
the universal function and the envelope correlation in the telegraph
approximation, (\ref{exp_tlgrph}).  The time scale of the exponential 
decay is $\mu\approx 50$ms for the slower varying stimulus, and 
$\mu\approx 20$ms for the faster varying one. In the case of the fast 
stimulus, the correlation is probably limited by the filtering processes
in the visual system, thus indicating that the time scale for these
processes is $\approx 20$ms. This is on the order of 
the behavioral time scale for changes in flight course in these flies,
found by Land and Collett (1974) to be $\approx 30$ms. 
{\narrowtext
\begin{figure}[h]
\vspace{0.2cm}
\begin{center}
\epsfxsize=8cm
\vspace*{-0.2cm}
\epsfbox{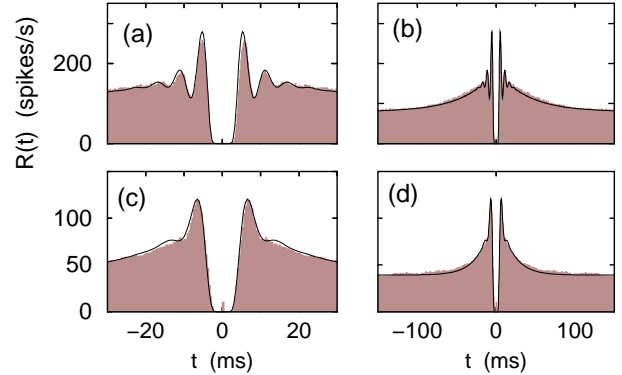}
\end{center}
\caption{
Conditional rates for spike trains, in the short time and long time
regimes. Data (gray histogram) are measured in experiments
where the visual stimulus was a pattern of light and dark bars moving
with a random time-dependent velocity. The velocity signal has
different
bandwidths, of 20 Hz (a,b) and 500 Hz (c,d). The theoretic formula
of Eq. (\ref{exp_tlgrph}) is presented as a solid black line.
}
\label{fig:9}
\end{figure}
}

\subsection{A Gaussian input signal}
If the incoming random signal is drawn from a Gaussian distribution,
one may calculate exactly the correlation function of a nonlinear
response $r(s)$. We focus on the simple case where the response is a step
function at zero, and the stimulus has zero mean, corresponding to a
coverage fraction of $\alpha=1/2$. In this case,
\begin{eqnarray}
\langle\Theta[s(t)]\Theta[s(0)]\rangle=\frac{1}{4}+\frac{1}{2\pi}
{\rm arcsin}(|c(t)|),
\end{eqnarray}
where $c(t)=\langle s(t)s(0)\rangle$ is the correlation function of the
Gaussian signal. In principle, one may calculate the correlation
function of a general nonlinear response $r(s)$ in the Gaussian case,
but for our purposes it is sufficient to consider the step response.
We expect that in an experiment where the driving stimulus
varies slowly, the effect of intermediate filters will be negligible and
one can take the motion on the screen to be essentially equivalent to
the stimulus $s(t)$ driving the neuron. 
Figure 10 shows the correlation function measured from an experiment
where the input signal was a slowly varying random function of time, with 
a Gaussian distribution. 
The correlation function as
calculated from the data is shown as a gray histogram, 
whereas the solid black line is given by 
\begin{eqnarray}
R(r)=
r R_U^{(r,D)}
\left[\frac{1}{4}+\frac{1}{2\pi}
{\rm arcsin}(|c(t)|)\right], \label{arcsin_sep}
\end{eqnarray}
where $c(t)$ was taken from the known distribution of the input signal.
The two fitting parameters are $r$, related to the global firing
rate, and $D$, the diffusion constant in the universal regime.

{\narrowtext
\begin{figure}[h]
\vspace{0.2cm}
\begin{center}
\epsfxsize=8cm
\vspace*{-0.2cm}
\epsfbox{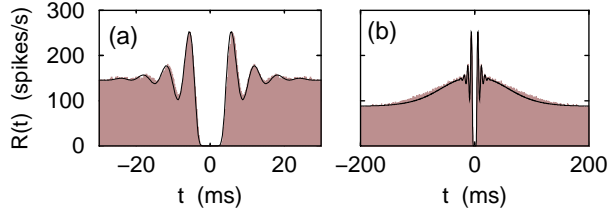}
\end{center}
\caption{
Conditional rate for spike trains, in the short time and long time
regimes. Data (gray histogram) are measured in experiments
where the visual stimulus was a pattern of light and dark bars moving
with a random time-dependent velocity, which is very slow and has a
Gaussian distribution.
The theoretic formula
of Eq. (\ref{arcsin_sep}) is presented as a solid black line.
}
\label{fig:10}
\end{figure}
}

\subsection{Linear rectifier approximation}
In analogy to Eq. (\ref{product1}), one expects that if the nonlinear response
of the neuron is $f(\cdot)$, the conditional rate will take the form
\be 
R(t)~~\approx~~ R_U^{(r,D)}~
\langle r[s(t)]r[s(0)]\rangle \label{product_gen}
\ee
\noindent 
where $r(s)$ is the firing rate as a function of stimulus, obtained by
averaging over noise only, and
$r,D$ are some effective global parameters representing
an average over different regimes where $s$ is almost constant.
In this section we show results from 
experiments where the response $r(s)$ cannot be approximated by a step
function,  
and apart from the direction selectivity the firing seems to 
follow the input stimulus linearly. A better approximation for the
response would therefore be a linear rectifier.
We used sine wave stimuli of different
amplitudes and periods.  Figure 11 shows
the firing rate averaged over noise, together with the stimulus,
for one such experiment. 

{\narrowtext
\begin{figure}[h]
\vspace{0.2cm}
\begin{center}
\epsfxsize=6cm
\vspace*{-0.2cm}
\epsfbox{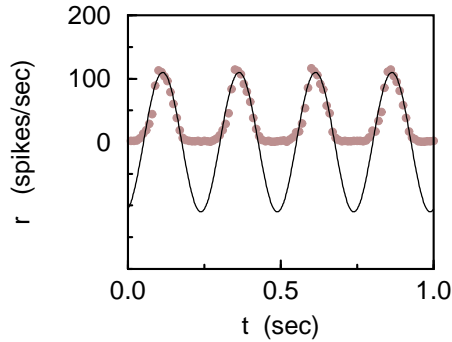}
\end{center}
\caption{
Stimulus and average response for a sine wave experiment. A random
pattern of dark and light horizontal lines was moved by a
sine wave velocity (solid black line), and this stimulus was
repeatedly
presented to the fly.
The average firing rate was calculated over the repeated presentations
as a function of time (circles).
The response of the
neuron is not saturated, and it follows closely the positive
part of th sine wave stimulus.
(Compare the saturated response in Fig. 1a).
}
\label{fig:11}
\end{figure}
}

We used the approximation 
\be
r(s)\approx s \Theta(s)
\ee
\noindent to evaluate the envelope of the conditional rate
$\langle r[s(t)]r[s(0)]\rangle$. For a sine wave stimulus of frequency
$\Omega$,
a straightforward calculation yields
\begin{equation}
\langle r[s(t)]r[s(0)]\rangle = 
\frac{1}{2}\cos(\Omega |t|)\Bigg[\frac{\pi}{\Omega}-|t|
\Bigg]
+\frac{1}{2\Omega}\sin(\Omega |t|).  \label{sin_env}
\end{equation}
\noindent The conditional rate is, then,  given by Eq. (\ref{product_gen}),
with Eq. (\ref{sin_env}) as the envelope.
If we use our knowledge about the frequency of the input
signal, we expect to get a description of the conditional rate
with only the universal fitting parameters $r$ and $D$, and no
additional parameters for the envelope.
Figure 12 shows two correlation functions for the sine wave experiments, in the
universal (a,c) and the stimulus-dependent (b,d) regimes.  No additional
parameters of the stimulus other than its frequency were used; the
solid black curve was obtained from Eq. (\ref{product_gen})
with the envelope given by (\ref{sin_env}).
The periodic nature of
the envelope is evident in the data, and the theoretic prediction
describes both the short range and the long range features well.

{\narrowtext
\begin{figure}[h]
\vspace{0.2cm}
\begin{center}
\epsfxsize=8cm
\vspace*{-0.2cm}
\epsfbox{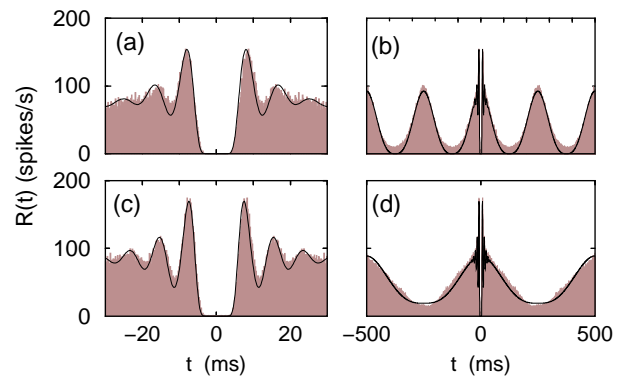}
\end{center}
\caption{
Conditional rates for spike trains in the short time and long time
regimes. Data (gray histograms) are measured in an experiment where
a pattern of dark and light bars were moved horizontally with a sine
wave velocity,
with periods of 0.25 sec (a,b) and 0.5 sec (c,d).
Theory (black solid lines) is obtained from Eq. (\ref{product_gen}),
with the
fitting parameters $r$ and $D$, and using the known frequency of the
stimulus.
}
\label{fig:12}
\end{figure}
}

\subsection{Irregularity and stimulus properties}
In this section, we showed that the conditional rate of the 
spike trains has approximately a product form. One term in the 
product is very similar to the conditional rate for a
constant stimulus, and describes the behavior on short times.
This term is {\em universal} in the sense that it depends on 
only two simple parameters, the global average rate
and the internal irregularity of firing. 
The other term carries information about temporal correlations in the
stimulus, and is a long-time envelope over the universal term. We have
shown how this envelope can be calculated in several cases, giving a very
good fit of the data. 

A naive quasi static application of the universal theory would tell us
that the short-time behavior is characterized by local values of the
parameters $r,D$, and that these change slowly as the external stimulus
varies slowly. This seems inconsistent with the short time behavior
exhibited by the data (Figures 8a, 10a, 10c). If the parameters would
change with the local changes of the external stimulus, the oscillatory
structure defined by a period of $r$ would be considerably
washed out when averaged over the different values of $s$. 
The data, however, show pronounced oscillations with a
well defined period. This is consistent with the notion that in a
changing environment, the system achieves a steady state with the
distribution as a whole, thus obtaining global values for the
statistical parameters of its firing. 

Further evidence for this
picture is given by Figure 13, where the correlation function is plotted
in normalized time units, and is normalized by the height of the first
peak. After rescaling of the two axes, there is only one dimensional
parameter describing the short time behavior: the irregularity $\gamma$.
Although the correlation functions were measured under different
experimental conditions of time varying stimuli, 
the curves are approximately overlapping in the short time
regime $(x<2)$. Thus,  under high signal-to-noise ratio,
where the frequency modulations are mainly induced by the stimulus and
not by the noise, the system tends to fix the internal irregularity
parameter $\gamma$ at some preferred value. This cannot be explained by
a simple quasi static behavior; it probably involves subtle adaptive
mechanisms of the neuron which try to maximize the dynamic range in each
stimulus ensemble. This subject is currently under further investigation
\cite{adapt}.

{\narrowtext
\begin{figure}[h]
\vspace{0.2cm}
\begin{center}
\epsfxsize=6cm
\vspace*{-0.2cm}
\epsfbox{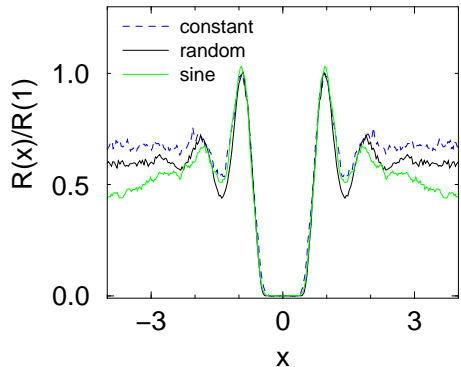}
\end{center}
\caption{
correlation functions from different experiments with different
time-varying stimuli, in dimensionless units. The overlap at short
times
indicates that the parameter left after rescaling, the internal
irregularity, is similar in all these cases.
}
\label{fig:13}
\end{figure}
}

\section{Discussion}

The statistical properties of spike trains
generated by a sensory neuron under various stimulation conditions were
considered.
Experiments were performed {\em in vivo} on a blowfly, where the visual
stimulus was well controlled, and spike trains  were measured
extracellularly from an identified motion sensor. This neuron is known
to respond to wide field horizontal motion.  

The main theoretical questions addressed were: (i) how does the
statistical behavior of spike trains {\em in vivo} relate to the
biophysics of the spike generation mechanism in the cell, and (ii) how 
can the effects of internal properties be separated from those induced
by the external sensory stimulus (or the input to the neuron).  These
questions were addressed within the framework of a model, which
describes the neuron as a nonlinear oscillator driven both by
noise and by an external stimulus. In general the effect of these two is
the same: to cause frequency modulations in the oscillator.
We considered here the case where the noise and the stimulus are very
different in their temporal characteristics, namely the noise is rapid
and the stimulus is slow, compared to the time scales typical of the
spiking. This separation of time scales
enables the theoretic analysis of the model,
which in turn provides an understanding of how the different factors 
are reflected in the statistical properties of the spike trains.

It was found that on the time scale of a few spikes, statistical
behavior is rather universal and can be described by a 
renewal process.
This process is derived from phase diffusion of a nonlinear
oscillator, and has special symmetry properties reflecting the
underlying oscillator. It is characterized by one dimensionless 
parameter which represents the
internal degree of irregularity of the process.  This parameter
interpolates between
highly regular and highly stochastic limits. The parametrizaion allows
spike trains from different parts of the visual system, and even from
different organisms, to be classified in a simple way.

On the time scale of many spikes, effects of the sensory stimulus become
important and are reflected in the form of a slowly varying envelope
which modulates the universal functions.
Using only simple features of the nonlinear neuronal response,
the theory provides quantitative predictions for the statistical
properties, which are found to be in very good agreement with our
measured data in both the short time and the long time regimes. 

Justification for assuming a separation of time scales between
noise and stimulus comes from the fact that signals in the
visual system are filtered, and therefore the effective signals reaching
an interneuron are relatively slow.  As often is the case in comparing
theory and experiment, we found that the agreement of the theoretical
predictions with the data extend to a regime where the
required separation of time scales is only marginally satisfied by the
experimental conditions. 

The understanding of how the stimulus is reflected in the statistical
properties of the spike train could be used ``backwards":
in cases where little is known about
the stimulus, analysis of the long time behavior of
the correlation function can give us information about the time scales involved
in this stimulus, with only a gross description of the neural response
(direction selectivity, degree of saturation). 
Possibly this understanding can be
extended to cross-correlation between several neurons; in that case the
separation between stimulus-induced and internal properties will be
important in assessing the connections between the neurons. This is subject for
future research.
\newline
\newline
\newline
\setcounter{equation}{0}
\def\theequation{A.\arabic{equation}}
{\large\bf Appendix A}\\ \\ 
In this Appendix we show that for a unidirectional random walk,
(a ``thermal ratchet'' walk),
the density of times between barrier crossing is 
 given by Eq.
(\ref{ps}), in the limit where many steps are needed to cross the barrier.

We consider the discrete case.
Let $x_i$ be non--negative random variables ($x_i\ge 0$), 
identically distributed
and independent, with $\bra x\ket=\mu$ and $\bra \delta x^2\ket
=\sigma^2$. Define the random variable of their sum as
\be
X_N= \sum_{i=1}^N x_i.
\ee
Let $C$ be a constant positive number, then for large enough $N$ one has
from the central limit theorem,
\begin{eqnarray}
{\rm Prob}\{X_N\le C\} &=& 
\frac{1}{\sqrt{2\pi N \sigma^2}} \int_{-\infty}^C \exp\{-
\frac{(X_N-N\mu)^2}{2N\sigma^2}\}\\
&=& 
\frac{1}{2}
\left[1+{\rm erf} \left(\frac{C-N\mu}{\sqrt{2N\sigma^2}}\right)\right]
\end{eqnarray}
where ${\rm erf}(x)=\frac{2}{\pi}\int_0^x e^{-t^2} dt$.
Now define $N_1$ to be the number of steps required to first reach the
barrier. Then from the non--negativity it follows that
\be
{\rm Prob}\{X_N\le C\} = {\rm Prob}\{N_1\ge N\} = 1-{\rm Prob}\{N_1\le N\}.
\ee
There are some normalization subtleties here, but let us imagine that 
the sample space is composed of trajectories with a finite number of 
steps $M$ which is much larger than the typical number needed to cross
threshold. Then, 
\be
{\rm Prob}\{N_1\le N\}~=~\frac{1}{2}
\left[1+{\rm erf} \left(\frac{N\mu-C}{\sqrt{2N\sigma^2}}\right)\right].
\ee
\newline
\newline
\newline
\setcounter{equation}{0}
\def\theequation{B.\arabic{equation}}
{\large\bf Appendix B}\\ \\ 
In this Appendix, we derive Eq. \ref{Rdim}. Starting from the definition,
\be
\langle\rho(t)\rho(0)\rangle ~~=~~ \Bigg\langle\sum_{k,l}\delta(\Phi(t)-k)
\delta(\Phi(0)-l)\dot{\Phi}(t)\dot{\Phi}(0)\Bigg\rangle.
\ee
The term at $t=0$ must be
taken care of separately: it corresponds to $k=k'$,
\begin{eqnarray}
&&\sum_k\Bigg\langle\delta\Bigg(\Phi(t)-k\Bigg)\delta\Bigg(\Phi(t)-\Phi(0)\Bigg)
\Bigg\rangle
\dot{\Phi}(t)\dot{\Phi}(0) \nonumber\\
&=&\sum_k
\Bigg\langle\delta\Bigg(\Phi(t)-k\Bigg)\frac{\delta(t)}{\dot{\Phi}(t)}
\dot{\Phi}(t)\dot{\Phi}(0)\Bigg\rangle \nonumber\\
&=&\delta(t') \sum_k\Bigg\langle\delta\Bigg(\Phi(t)-k\Bigg){\dot{\Phi}(t)}
\Bigg\rangle \nonumber\\
&=&r~\delta(t),
\end{eqnarray}
\noindent As expected for a point process of average rate $r$. For times 
$t \neq 0$, we calculate the conditional rate
$R(t)$, the
probability of firing at some time $t$ given a spike at time $0$: 
\begin{eqnarray}
R(t)&=& \sum_{k\neq 0}\langle\delta(\Phi(t)-k){\dot{\Phi}(t)}
\rangle\nonumber\\
&=& \sum_{k\neq 0} \frac{d}{dt} \langle\Theta(\Phi(t)-k)\rangle.
\end{eqnarray}
Using the integral representation of the Theta function, we write
\begin{eqnarray}
R(t)=\sum_{k\neq 0} \frac{d}{dt}\int_{-\infty}^{\infty}
\frac{dp}{2\pi i p} e^{-i2\pi k} 
\langle e^{i2\pi \Phi(t)}\rangle. \nonumber\\
\end{eqnarray}
As before, we use the Gaussian approximation for $\Phi(t)$, Eq. 
(\ref{Gauss_approx}), which is
justified for times $t$ satisfying $\tau_n\ll t$:
\begin{eqnarray}
R(t)&=&\sum_{k\neq 0} \frac{d}{dt}\int_{-\infty}^{\infty}
\frac{dp}{2\pi i p} 
e^{i2\pi p(rt-k)-2\pi^2 p^2 Dt}\nonumber \\
&=&\sum_{k\neq 0} \frac{d}{dt}\int_{-\infty}^{\infty} dp
\int^{\nu}dy e^{i2\pi y-2pi^2 p^2 Dt}
\end{eqnarray}
where $\nu=rt-k$. Doing first the integral $dp$, we find
\begin{eqnarray}
R(t)&=&\sum_{k\neq 0} \frac{d}{dt} \int^{\nu}\frac{dy}
{\sqrt{2\pi Dt}} ~e^{y^2/2Dt} \nonumber\\
&=&\sum_{k\neq 0} \frac{d}{dt} \left[\frac{1}{2}
\mbox{erf}\left(\frac{\nu}{\sqrt{2Dt}}\right) + \mbox{const}\right]
\nonumber \\
&=& r \sum_{k\neq 0} \frac{t+k/r}{\sqrt{8\pi Dt^3}} e^{(rt-k)^2/2Dt}.
\end{eqnarray}
\newline
\newline
\newline
\setcounter{equation}{0}
\def\theequation{C.\arabic{equation}}
{\large\bf Appendix C}\\ \\ 
Here we discuss the properties of the number variable, which
counts the number of spikes in the window
$[0:t]$. One must specify how the point $t=0$ is chosen,
and there are two natural choices: (i) start counting at a spike,
and (ii) start counting at a random point in the spike train. 
The second choice of a random 
origin is the more commonly used. In this case, it is convenient to
define an auxiliary variable $\varphi=\Phi(t_1)$, the phase of the
integrator $\Phi(t)$ at the time of the first spike in the window. This 
variable is uniformly distributed in $[0:1]$, and with this we have
\begin{eqnarray}
N(t)&=&\mbox{Int}[\Phi\!(t)\!+\!1\!-\!\varphi]\nonumber\\
=\Phi\!(t)\!&+&\!1\!-\!\varphi +\sum_{m=1}^{\infty}
\frac{1}{\pi m} \sin[2\pi m(\Phi\!(t)\!+\!1\!-\!\varphi] - \frac{1}{2}.
\end{eqnarray}
Note that the constant $-\frac{1}{2}$ 
ensures that $N(0)=0$. Since $\varphi$ only
depends on the choice of origin, it is independent of $\Phi(t)$, and
therefore 
\begin{eqnarray}
\langle N(t)\rangle &=&\langle \Phi(t)\rangle 
+\frac{1}{2}-\langle\varphi\rangle +\sum_{m\neq 0}\frac{1}{2\pi i m} 
\langle e^{i2\pi m[\Phi(t)\!+\!1\!-\!\varphi]}\rangle   \nonumber \\
&=& rt+\sum_{m\neq 0}\frac{1}{2\pi i m}\langle e^{[i2\pi m[\Phi(t)\!+\!1]}
\rangle\langle e^{-i2\pi m \varphi}\rangle \nonumber \\
&=& rt.
\end{eqnarray}
Now to calculate the number variance we define the fluctuation
\be
\delta N(t) = \delta \Phi(t)+(\varphi-\frac{1}{2}) +\sum_{m\neq 0}
\frac{1}{2\pi im} e^{i2\pi m[\Phi(t)+1]}
\ee
and average its square:
\begin{eqnarray}
\sigma^2_N&=&\langle \delta N(t)^2\rangle=
\langle \delta \Phi\!(t)^2\rangle+\langle(\frac{1}{2}\!-\!\varphi)^2\rangle
\nonumber\\
+&2&\langle \delta \Phi\!(t)(\frac{1}{2}\!-\!\varphi)\rangle
+\langle 2(\frac{1}{2}\!-\!\varphi)\sum_{m\neq 0}\frac{1}{2\pi im}
e^{[i2\pi m[\Phi\!(t)\!+\!1\!-\!\varphi]}\rangle \nonumber\\
&+&\sum_{m,m'\neq 0}\frac{1}{(2\pi i)^2 m m'}
\langle e^{i2\pi(m+m')[\Phi(t)+1-\varphi]}\rangle. \label{av-sq}
\end{eqnarray}
By Eq. (\ref{cumulants}), $\langle \delta \Phi(t)^2\rangle\!=\!Dt$;
averaging the second term over $\varphi$ gives $\frac{1}{12}$.
Due to the independence of $\varphi$ and $\Phi(t)$, 
the first cross-term vanishes 
while the second cross-term decouples into
\be
-2\sum_{m\neq 0}\frac{1}{2\pi im}\langle e^{[2\pi i m[\Phi(t)\!+\!1]}\rangle
\langle \varphi e^{-2\pi i m\varphi}\rangle.
\ee
Performing the average over $\varphi$, 
\be
\langle \varphi e^{-i2\pi m\varphi}\rangle = \frac{d}{d\mu}
\frac{1}{2\pi i m}\langle e^{-2\pi im\varphi
\mu}\rangle\Bigg|_{\mu\!=\!1}=\frac{-1}{2\pi im},
\ee
and we have for the second cross-term
\be
\sum_{m\!=\!1}^{\infty} \frac{1}{(\pi m)^2}\cos(2\pi mrt)e^{-2\pi^2 m^2
Dt},
\ee
where the Gaussian approximation was used when averaging over $\Phi(t)$.
In the last double sum of Eq. (\ref{av-sq}), 
all terms vanish by averaging over
$\varphi$ except for the terms $m+m'=0$, which gives
\be
\sum_{m\!\neq\!0}\frac{1}{(2\pi m)^2}=\frac{1}{12}.
\ee
Adding the terms together we find 
\be
\sigma^2_N=Dt+\frac{1}{6} +\sum_{m\!=\!1}^{\infty}\frac{1}{(\pi m)^2}\cos(2\pi
mrt)e^{-2\pi^2 m^2 Dt}
\ee
which is the same as Eq. (\ref{NV}).
\newline
\newline
\newline
\setcounter{equation}{0}
\def\theequation{D.\arabic{equation}}
{\large\bf Appendix D}\\ \\ 
In this Appendix we derive Eq. (\ref{exp_tlgrph}) for the correlation
function of the envelope of the spike train, in the telegraph
approximation. The envelope $\rho^E(t)$ is composed of a train of
characteristic window functions,
\begin{eqnarray}
\langle\rho^E(t)\rangle = \sum_k \chi\Bigg(T_k\!-\!\frac{\Delta_k}{2},T_k
\!+\!\frac{\Delta_k}{2}\Bigg)
\end{eqnarray}
\noindent as illustrated in Fig. 14. 
This function has a Fourier transform
\begin{eqnarray}
\langle\rho^E(\omega)\rangle &=&
\int_{-\infty}^{\infty}\rho^E(t)e^{-i\omega t}\nonumber\\
&=& \sum_k e^{-i\omega T_k} \frac{\sin(\omega \Delta_k/2)}{(\omega/2)}
+\alpha\delta(\omega),
\end{eqnarray}
where $T_k$ denote the middle points of the positive signal regions.

{\narrowtext
\begin{figure}[h]
\vspace{0.2cm}
\begin{center}
\epsfxsize=8cm
\vspace*{-0.2cm}
\epsfbox{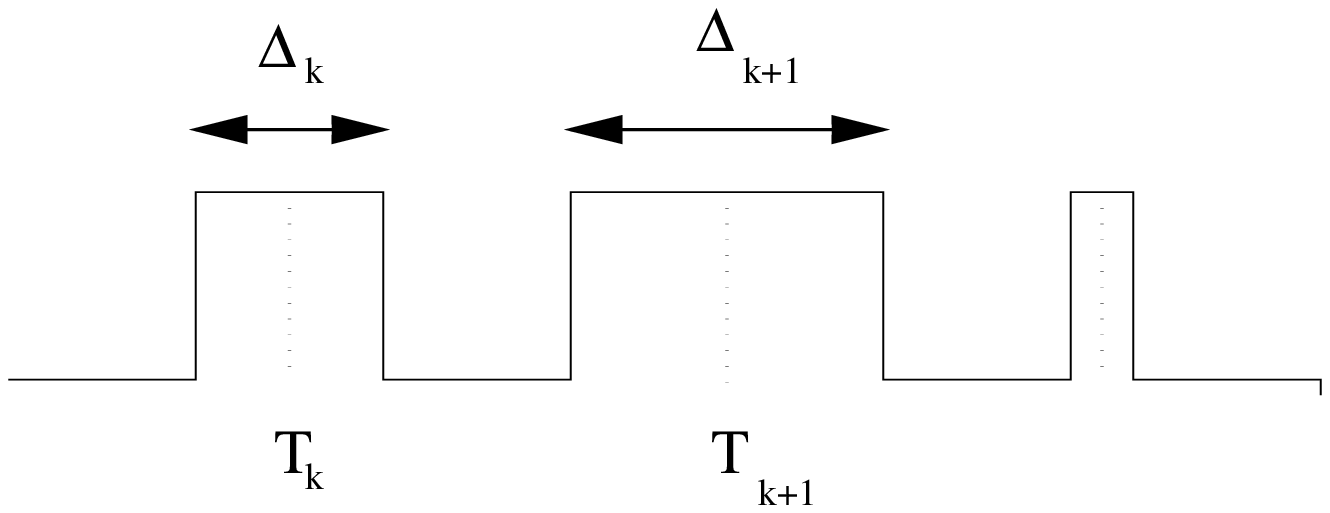}
\end{center}
\caption{
}
\label{fig:14}
\end{figure}
}

These positive regions are assumed to be
distributed over the time axis independently with a density $\beta$ per unit
time, and with an average length of $\langle\Delta\rangle=\mu$. Averaging
in the frequency domain gives
\be
\langle|\rho^E(\omega)|^2\rangle=\beta \frac{\langle\sin^2(\omega
\Delta/2)\rangle}{(\omega/2)^2}
+ (\beta\mu)^2 \delta(\omega).
\ee
In the time domain this expression transform to
\begin{eqnarray}
\langle\rho^E(t)\rho^E(0)\rangle &=& (\beta\mu)^2 \\
\nonumber 
&+&\beta \int_0^{\infty}p(\Delta)d\Delta\int_{-\infty}^{\infty}
\frac{d\omega}{2\pi} \frac{\sin^2(\omega\Delta/2)}{(\omega/2)^2}
e^{i\omega t}.
\end{eqnarray}
\noindent The last tern in this expression is the Fourier transform of
a product of two sinc function, which is the convolution of two square
windows. This convolution has the form
\begin{eqnarray} 
\chi*\chi = \Bigg\{ \begin{array}{l}
0~~~~~, \hfill |t|>\Delta\\
\Delta-|t|,\hfill |t|\leq \Delta
\end{array}
\end{eqnarray}
which is equivalent to Eq. (\ref{env_corr}).  \\ \\ \\ 

\noindent{\bf Acknowledgments}\\
Many thanks to G. Lewen, for preparing the experiments,
and to N. Tishby and A. Schweitzer for comments.


\end{multicols}
\end{document}